\documentclass[12pt]{article}

\usepackage{graphicx}
\begin{document}

\rightline{LYCEN 9919}

\rightline{March 25-1999}

\bigskip

\begin{center}

{\bf {\Large {HOW FAST DO SMALL $x$ STRUCTURE FUNCTIONS
RISE~? \ \ A COMPARATIVE ANALYSIS}}}

\bigskip

{\large {\
P. Desgrolard (\footnote{{ E-mail: desgrolard@ipnl.in2p3.fr}}),
L. Jenkovszky (\footnote{{ E-mail: jenk@bitp.kiev.ua}}),
A. Lengyel (\footnote{{ E-mail: sasha@len.uzhgorod.ua}}),
F. Paccanoni(\footnote{{ E-mail: paccanoni@pd.infn.it}})
}}

\end{center}

\bigskip

($^1$) {\it Institut de Physique Nucl\'eaire de Lyon, IN2P3-CNRS et
Universit\'e Claude Bernard, 43 boulevard du 11 novembre 1918, F-69622
Villeurbanne Cedex, France}

($^2$) {\it Bogolyubov Institute for Theoretical Physics, National Academy
of Sciences of Ukraine, 252143 Kiev-143, Metrologicheskaja 14b, Ukraine}

($^3$) {\it Institute of Electron Physics, National Academy of Sciences of
Ukraine, 294015 Uzhgorod-015, Universitetska 21, Ukraine}

($^4$) {\it Dipartimento di Fisica, Universit\'a di Padova, Istituto
Nazionale di Fisica Nucleare, Sezione di Padova, via F. Marzolo,I-35131,
Padova, Italy}

\bigskip
\bigskip
\noindent
{\bf {\large Abstract}} We parametrize the small $x$, singlet component of
the proton structure function $F_2$ by powers and logarithms of ${\frac{1}{x}%
}$ for discrete values of $Q^2$ between 0.2 and 2000 GeV$^2$, and compare
these parametrizations by applying the criterion of minimal $\chi^2$. The
obtained values of the fitted parameters may be used to study the evolution
of $F_2$ in $Q^2$ and/or in discriminating between dynamical models. A
slowing-down in the increase of $F_2$ towards highest available values of $%
Q^2$ is revealed. The effect is quantified in terms of the derivative ${%
\frac{d\ell n F_2(x,Q^2)}{{d\ell n (1/x)}}}$.

\bigskip

\section{Introduction}

It is customary (for a recent review see, e.g.~\cite{CDD}) to parametrize
the small $x(x<0.1)$ behaviour of the proton structure function (SF) by a
power-like growth
\begin{equation}
F_2(x,Q^2)\sim a\left( Q^2\right) x^{-\lambda (Q^2)}\ ,
\end{equation}
with the $Q^2$ dependent "effective" power $\lambda (Q^2)$ generally
interpreted as the Pomeron intercept $-1$, rising from about $0.1$ to about $%
0.4$ between the smallest and largest values of $Q^2$ measured at HERA.

In recent papers~\cite{dl,cdl} a "hard" Pomeron term, with $\epsilon
_0=0.418 $, besides the "soft" one, with $\epsilon _1=0.0808$ and a
subleading "Reggeon" with $\epsilon_2=-0.4525$ (all $Q^2$ independent) was
introduced in the SF
\begin{equation}
F_2(x,Q^2)\sim \sum_{i=0}^2a_i\left( Q^2\right) x^{-\epsilon _i}\ .
\end{equation}
In our opinion, there is only one Pomeron in
the nature~\cite{BGJPP}: nevertheless, for the sake of completeness, we have 
included in our analysis the above parametrization as well. 

\noindent
Notice that with an increasing number of contributions to $F_2$ the values
of some of the powers (related to the intercepts of relevant trajectories)
must be fixed in some way since the number of the small $x$ data is not
sufficient to determine unambiguously their values from the fits.

Alternative logarithmic parametrizations 
\begin{equation}
F_2(x,Q^2)\sim \sum_{i=0}^2b_i\left( Q^2\right) \ell n ^{_i}\left( \frac
1x\right) \
\end{equation}
exist and are claimed~\cite{log} to be equally efficient.

Notice also that in expressions (2),(3), contrary to (1), the $Q^2$
dependence factorizes in each individual term (Reggeon) - a typical feature
of the Regge pole theory - (for more details see \cite{BGJPP}).

\noindent
It should be noted that each term in the simple
parametrizations of the type (2) and (3) may be associated with the
$Q^2$ independent Pomeron trajectories with relevant factorized
$Q^2$ dependent residuae. Formally, it is not compatible with the
GLAP evolution equation, by which the variation (evolution) with
$Q^2$ modifies also the $x$ dependence of the SF (although in a
limited range, approximate "selfconsistent" solutions, stable with
respect to a logarithmic behaviour are known \cite{JLP} to exist).
In any case, since we are fitting the SF to fixed values (bins) of
$Q^2$, our parametization does not depend directly  on the $Q^2$
evolution. Moreover, since the onset and range of the perturbative
GLAP evolution is not known {\it a priori}, our "data" may be used as a
test for it. 

In this paper we present the results of a comparative analysis of these two
types of parametrizations~: power-like and logarithmic. To avoid theoretical
bias, we do not constrain the $Q^2$ dependence by any particular model,
instead we take the experimental value in a parametric way.

\noindent
The range of variables and the set of experimental points are, of course,
the same in both cases. As a by-product, the parameters obtained in this way
may be used as "experimental data" in future calculations of the GLAP or
BFKL evolution. The present study is an extension of a preliminary 
analysis~\cite{JLP}.

\noindent
It is generally believed that, at small $x$, the singlet SF increases
monotonically, indefinitely, accelerating towards larger $Q^2$ (the Pomeron
becomes more "perturbative"). This phenomenon is usually quantified by means
of the derivative $\partial \ell n {F_2/\partial {\ln (1/x)}}$, which in the
simple case of $F_2\sim x^{-\lambda }$ ($\lambda$ being $x$
independent), is identical with the effective power $\lambda $
(otherwise it is not). We found evidence against this monotonic
trend: moreover, we show that, at the highest $Q^2$, the rise of
$F_2$ starts slowing down.

\bigskip

\section{Analysis of the structure function}

\subsection{Small $x\ (< 0.05)$}

The following forms of the small $x$ singlet component ($S,0$) of the SF are
compared for $x\ < x_c$ and for each experimental $Q^2_i$ bin~:

\smallskip

A. Power-like
\begin{equation}
F_2^{S,0}(x,Q_i^2)=a(Q_i^2)({\frac 1x})^{\lambda (Q_i^2)},
\end{equation}
and
\begin{equation}
F_2^{S,0}(x,Q_i^2)=a_0(Q_i^2)\ ({\frac 1x})^{\epsilon _0}\ +a_1(Q_i^2)\ ({\frac
1x)}^{\epsilon _1},
\end{equation}
where the exponents $\epsilon_0,\epsilon_1$ are fixed in accordance with~%
\cite{dl,cdl}.
\smallskip

B. Logarithmic
\begin{equation}
F_2^{S,0}(x,Q_i^2)=b_0(Q_i^2)+b_1(Q_i^2)\ell n({\frac 1x})\ ,
\end{equation}
\begin{equation}
F_2^{S,0}(x,Q_i^2)=b_0(Q_i^2)+b_2(Q_i^2)\ell n^2({\frac 1x})
\end{equation}
and the combination of the two
\begin{equation}
F_2^{S,0}(x,Q_i^2)=b_0(Q_i^2)+b_1(Q_i^2)\ell n({\frac 1x})\ +b_2(Q_i^2)\ell
n^2({\frac 1x})\ .
\end{equation}

\smallskip

In these equations, $a(Q_i^2)$, $a_{0,1}(Q_i^2)$, $b_{0,1,2}(Q_i^2)$ and $%
\lambda (Q_i^2)$ are parameters fitted to each $i^{{\rm th}}$ $Q^2$ bin.
More precisely, the free parameters are $a$ and $\lambda $ for (4), $a_0$
and $a_1$ for (5), $b_0$ and $b_1$ for (6), $b_0$ and $b_2$ for (7), $b_0,$ $%
b_1$ and $b_2$ for (8).

The choice of the cut $x_c$ is obviously crucial, but subjective. Balancing
between $x$ small enough, to minimize the large $x$ effects, and $x$ large
enough to include as many data points as possible, we tentatively set, like
in~\cite{JLP} $x_c=0.05$ as a compromise solution.

Since one can be never sure of the choice of the boundary below which the
non-singlet contribution ($nS,0$) may be neglected \footnote{%
The ratio of the non-singlet contribution to the singlet one was calculated
in \cite{BDI} and was shown to drop below 10\% around $x=10^{-3}$,
tending to decrease with increasing virtualities $Q^2$.}, we performed
additional fits with the non singlet contribution included
\begin{equation}
F_2^{nS,0}(x,Q_i^2)=a_f\left( Q_i^2\right) x^{1-\alpha _f}\ ,
\end{equation}
with the intercept fixed as in~\cite{kaidalov}, $\alpha _f=0.415$ ({\it i.e.}
only one free parameter, namely $a_f$ is added).


\subsection{Extension to all $x\ (< 1.0)$}

To ensure that our fits do not depend on the choice of the cut $x_c,$ we
extend the previous analysis to larger values of $x$ with relevant
modifications of the SF. Namely, we multiply the singlet and
subsequently the non singlet contributions by appropriate large $x$
factors~\cite {kaidalov}. The resulting SF becomes \begin{equation}
F_2(x,Q_i^2)=F_2^{S,0}(x,Q_i^2)(1-x)^{n\left( Q_i^2\right)
+4}+F_2^{nS,0}(x,Q_i^2)(1-x)^{n\left( Q_i^2\right) }\ ,
\end{equation}
where $F_2^{S,0}$ runs over all the cases considered in the previous
section, and the exponent $n\left( Q_i^2\right)$ is either that of~\cite
{kaidalov}
\begin{equation}
n\left( Q_i^2\right) =\frac 32\left( 1+\frac{Q_i^2}{Q_i^2+c} \right)\ ,\
{\rm with}\ c=3.5489\, {\rm GeV}^2\ ,%
\end{equation}
or is fitted to the data for each $Q_i^2$ value (see below).

\bigskip

\section{Discussion of the results}

\subsection{Structure function}

We made two kinds of fits, one restricted to small $x$ only, ($%
x\,<\,x_c=0.05 $), the other one including large $x$ as well. In the first
case ($x\,<\,x_c$) the experimental data are from \cite{HERA1}. Altogether
43 representative $Q^2$ values were selected to cover the interval $%
[0.2,1200]$ GeV$^2$ and $x\in [2.10^{-6},x_c]$. Including more (or all
available) data points had little effect on the resulting trend of the
results. The relevant values of $\chi ^2$, with and without the non-
singlet term, are given in Table 1. Notice, that we use the
definition
\begin{equation} <\chi ^2/{\rm
dof}>=\frac{\sum_{i=1}^{N_{bin}}\left( \frac{\chi _i^2}{n_{{%
data}_i}-m_{para}}\right) }{N_{bin}}\ ,
\end{equation}
where each $Q^2_i$ bin out of a total of $N$ bins contains $ n$ data
points and gives a resulting contribution to $\chi _i^2$ in fitting
eqs. (4)-(8), each containing $m$ parameters.

\medskip

{\bf Table 1}. Results of the fits without or with non-singlet term (9)
for small $x\,(<0.05)$. The total number of experimental points is 508.

\medskip

\begin {tabular}{cccccc}
\hline
Version & Power & Power & Logarithm & Logarithm & Logarithm \\
\hline\hline
Eq.     & (4)   &  (5)  &    (6)    &     (7)   &   (8)     \\
Nb. of parameters & 2 & 2 & 2 & 2 & 3 \\
$\chi ^2$  & 282 & 262 & 463 & 303 & 231 \\
$<\chi^2/{\rm dof}> $ & 0.68 & 0.62 & 1.04 & 0.70 & 0.61\\
\hline
Eqs.     & (4,9)   &  (5,9)  &    (6,9)    &     (7,9)   &   -   \\
Nb. of parameters & 3 & 3 & 3 & 3 & - \\
$\chi ^2$  & 247 & 225 & 253 & 234 & -\\
$\chi ^2/{\rm dof}$ & 0.67 & 0.60 & 0.67 & 0.62 & - \\
\hline
\end {tabular}

\medskip

\noindent
Notice that in performing the small $x$ fit we profited from a large set of
available data, while in the large $x$ extension a representative set of 30 $%
Q^2$ bins ($Q^2\in [1.5, 2000]$ GeV$^2$) was used. The data are from \cite
{HERA1,NMC}. The relevant $\chi ^2$ values are shown in Table 2. Two options
are presented~: the first one relies entirely on the extension by~\cite
{kaidalov}, in the second one the exponent $n(Q^2)$ (see (10)) is fitted for
each $Q^2$ bin.

            \vfill\eject

\medskip

{\bf Table 2}. Results of the fits for all $x\ (<1.0)$, when the parameters of
the large $x$ extension $n$ is chosen as in~\cite{kaidalov} or fitted.  
The total number of experimental points is 545.

\medskip

\begin {tabular}{cccccc}
\hline
Version & Power & Power & Logarithm & Logarithm & Logarithm \\
\hline\hline
Eqs. & (4,9,10,11)&(5,9,10,11)&(6,9,10,11)&(7,9,10,11)&(8,9,10,11)\\
Nb. of parameters & 3 & 3 & 3 & 3 & 4 \\
$\chi ^2$   & 368 & 371 & 894 & 399 & 319 \\
$<\chi^2/{\rm dof}> $ & 0.79 & 0.79 & 1.74 & 0.85 & 0.78 \\
\hline
Eqs.     & (4,9,10)   &  (5,9,10)  &  (6,9,10) & (7,9,10)  &   -   \\
Nb. of parameters & 4 & 4 & 4 & 4 & - \\
$\chi ^2$   & 321 & 317 & 541 & 329 -\\
$\chi ^2/{\rm dof}$ & 0.76 & 0.75 & 1.25 & 0.780& - \\
\hline
\end {tabular}

\medskip

The $Q^2$ dependence of the parameters was shown in Figs.~1-3.
We exposed the most representative results from the small $x$ fit that may
clarify asymptotic trends in the behaviour of the singlet SF (see~\cite{JLP}
and the
following discussion of the results). As already explained, the large $x$
extension was intended merely to support the small $x$ results.

\begin{center}
\includegraphics*[scale=0.9]{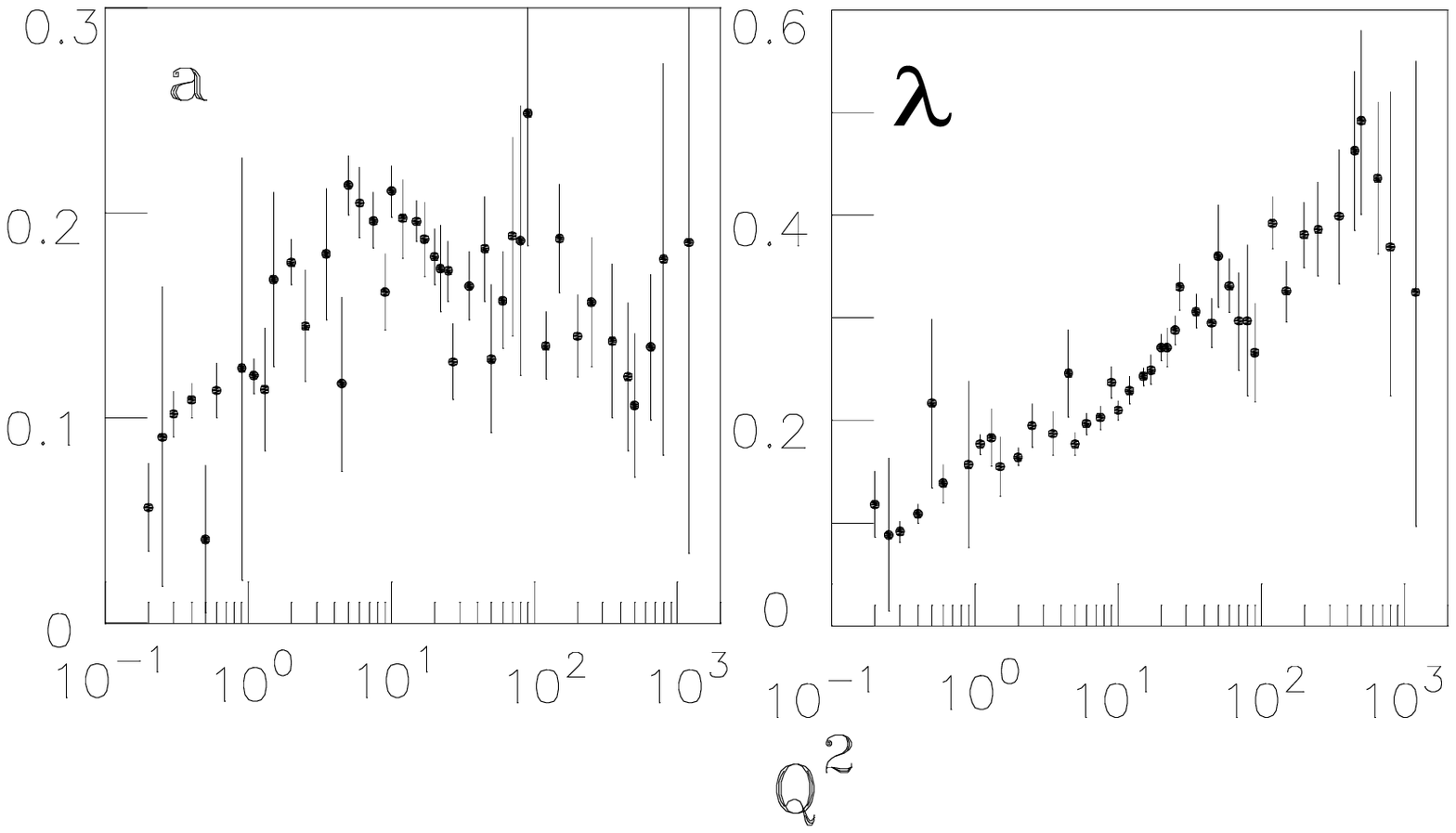}
\end{center}

Fig.~1. Results of our analysis for $a\left( Q_i^2\right) $ and $\lambda
\left( Q_i^2\right) $ entering in the parametrization (4)~: $F_2^{S,0}=a({%
\frac 1x})^\lambda $ of the small $x$ structure function ($x<x_c=0.05$);
they are fitted to the discrete values of $Q^2$ data from
\cite{HERA1}; $Q^2$ is in GeV$^2$, the error bars are produced from
the minimization program "Minuit".

\begin{center}
\includegraphics*[scale=0.85]{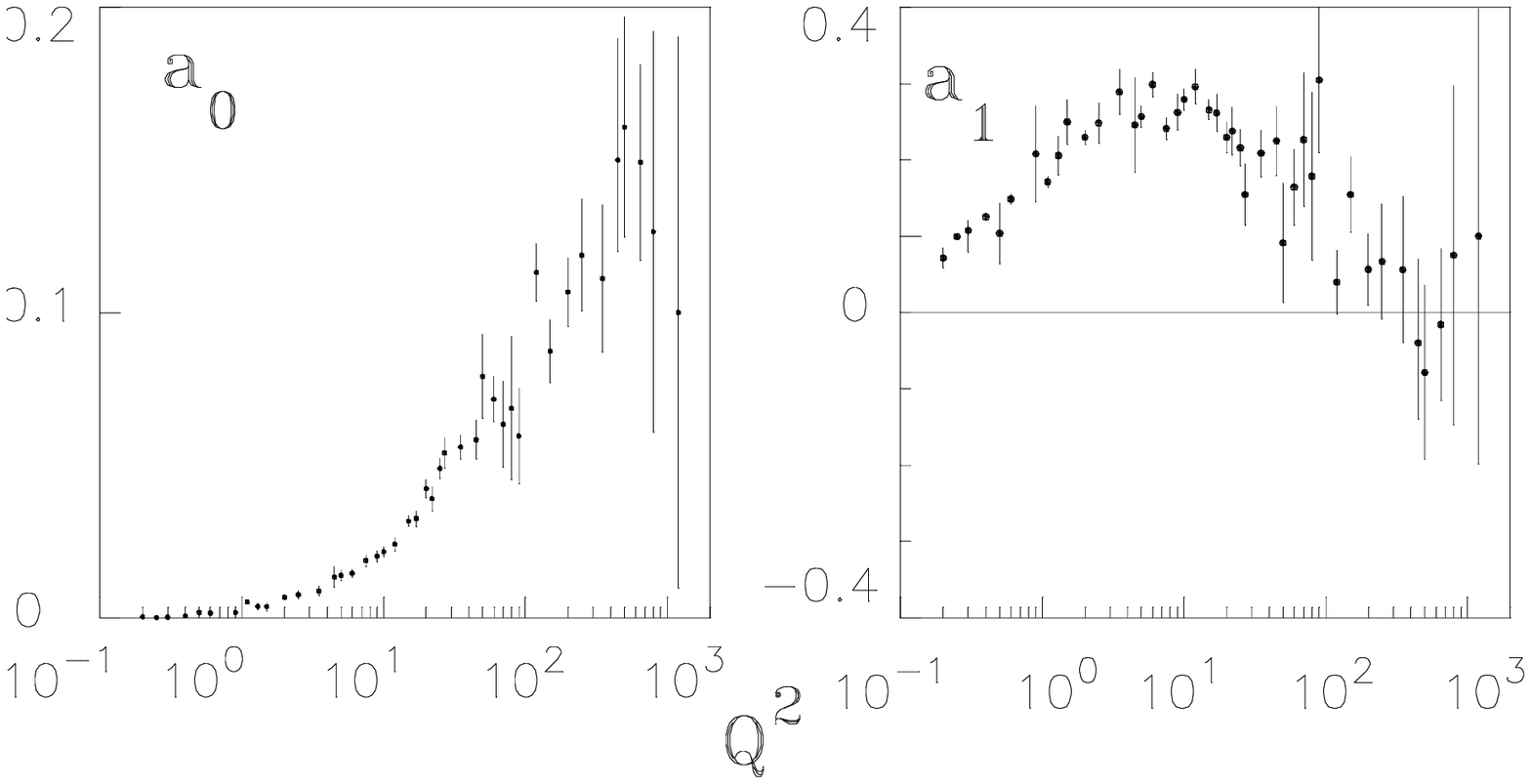}
\end{center}

Fig.~2. Same as Fig~.1 for $a_0\left( Q_i^2\right)$, $a_1\left( Q_i^2\right)$
and parametrization (5)~: 

$F_2=a_0({\frac 1x})^{0.418}\ +a_1({\frac 1x)}^{0.0808}$.

\begin{center}
\includegraphics*[scale=0.95]{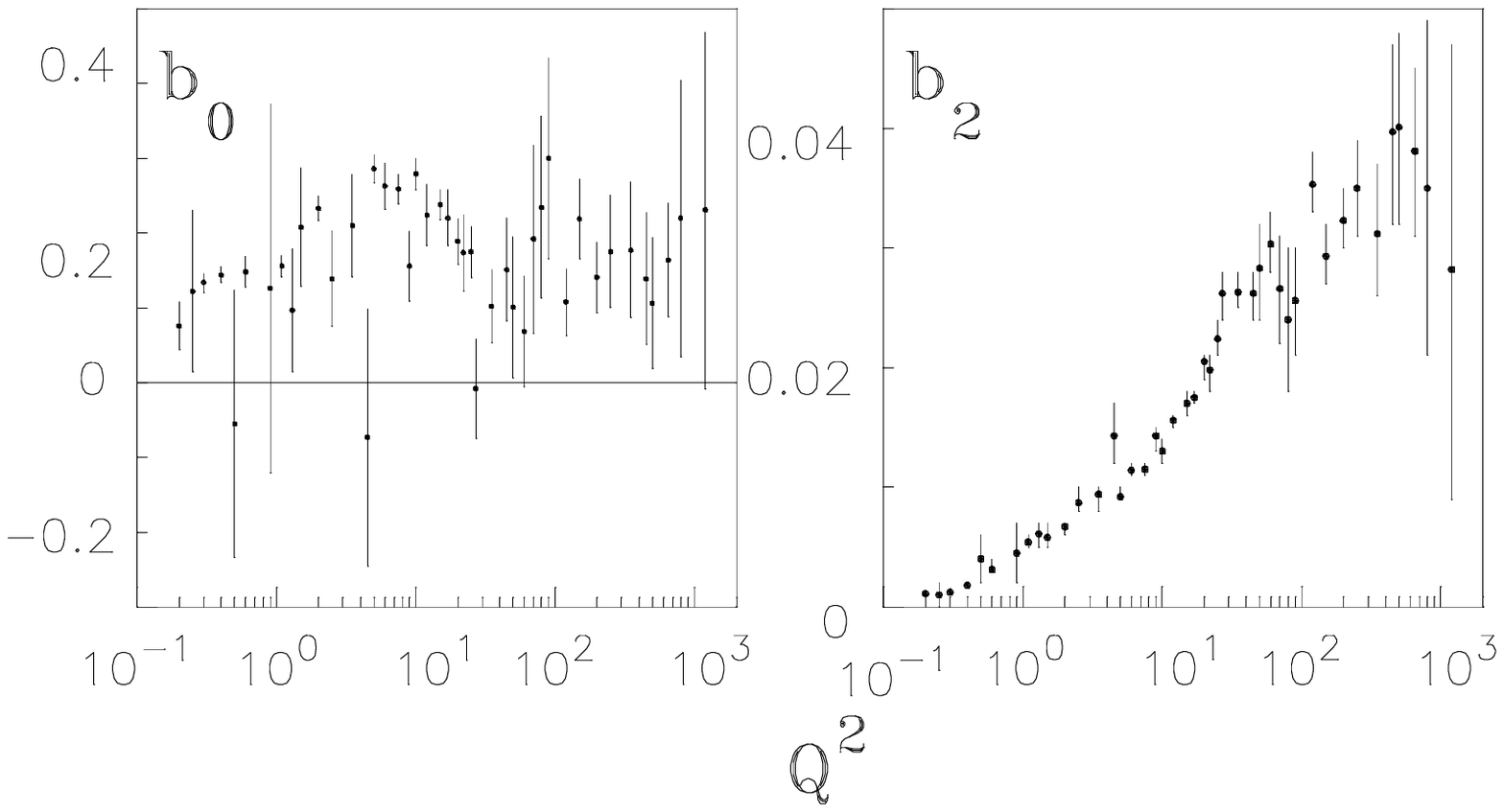}
\end{center}

Fig.~3. Same as Fig~.1 for $b_0\left( Q_i^2\right) $ and $b_2\left(
Q_i^2\right) $ and parametrization (7)~: 

$F_2^{S,0}=b_0+b_2\ell n^2({\frac 1x})$.

\bigskip

The following comments are in order~:

All the parametrizations (4)-(8), except (6), result in roughly equal
quality fits. We may rule out the parametrization (6) giving the poorest
(as expected) agreement with the data; so we do for
the least economic (largest number of the free parameters),
parametrization (8) (not shown in the figures).

The best results are achieved for the parametrization (8), giving the best
value of the total $\chi^2$. Although fit (8) contains an extra free
parameter with respect to the rest, the $\chi ^2/{\rm dof}$ value is,
nevertheless, better than in other variants (4)-(7). Notice that (8) leads
to alternating signs of the coefficients. Such an effect has been observed
earlier in a fit to hadronic total sections~\cite{IL}; it resembles the
first few terms in an expansion of the supercritical Pomeron in an alternate
series of logarithms.

\subsection{$x$-slope}

A clear indicator measuring the rate of increase of $F_2$ is its logarithmic
derivative or the $x-$ slope
\begin{equation}
B_x(x,Q_i^2)={\frac{\partial \ell nF_2 ( x,Q_i^2) }{\partial \ell n{\frac 1x}%
}}
\end{equation}
identical with the effective power $\lambda (Q_i^2)$ in the case of a single
power term, $x$ independent as in (1) \footnote{%
Note that by factorization, the intercept is $Q^2$ independent; that is why
(1) is an "effective" Regge pole contribution rather than a genuine Pomeron
\cite{BGJPP}.} (for an example of utilization of this derivative, 
see~\cite{schof}).

The $x$-slope $B_x$ is a function depending on two variables $x$ and $Q^2$,
which in principle are independent, although correlated by a
kinematical constraint: $y\le 1$, which at HERA energy becomes
$Q^2$ (in GeV$^2$) $< 9.10^4 x$.
The derivative can be calculated either
analytically, if the SF is parametrized explicitly, or numerically by
calculating the finite difference within certain intervals $<x>$. If a given
parametrization fits the data well, then the analytical differenciation has
a chance to reflect the slope, although it will not be model-independent. By
calculating the slopes of finite bins, we have a better chance to be
model-independent, although the result may depend on the width of the
chosen bins.

For the parametrization (4) $B_x=\lambda$
is already shown in Fig.~1. We show in Fig.~4 the results of 
our analysis corresponding to the other representative
cases (5),(7), the coefficients of which are exhibited above.
\begin{center}
\includegraphics*[scale=0.8]{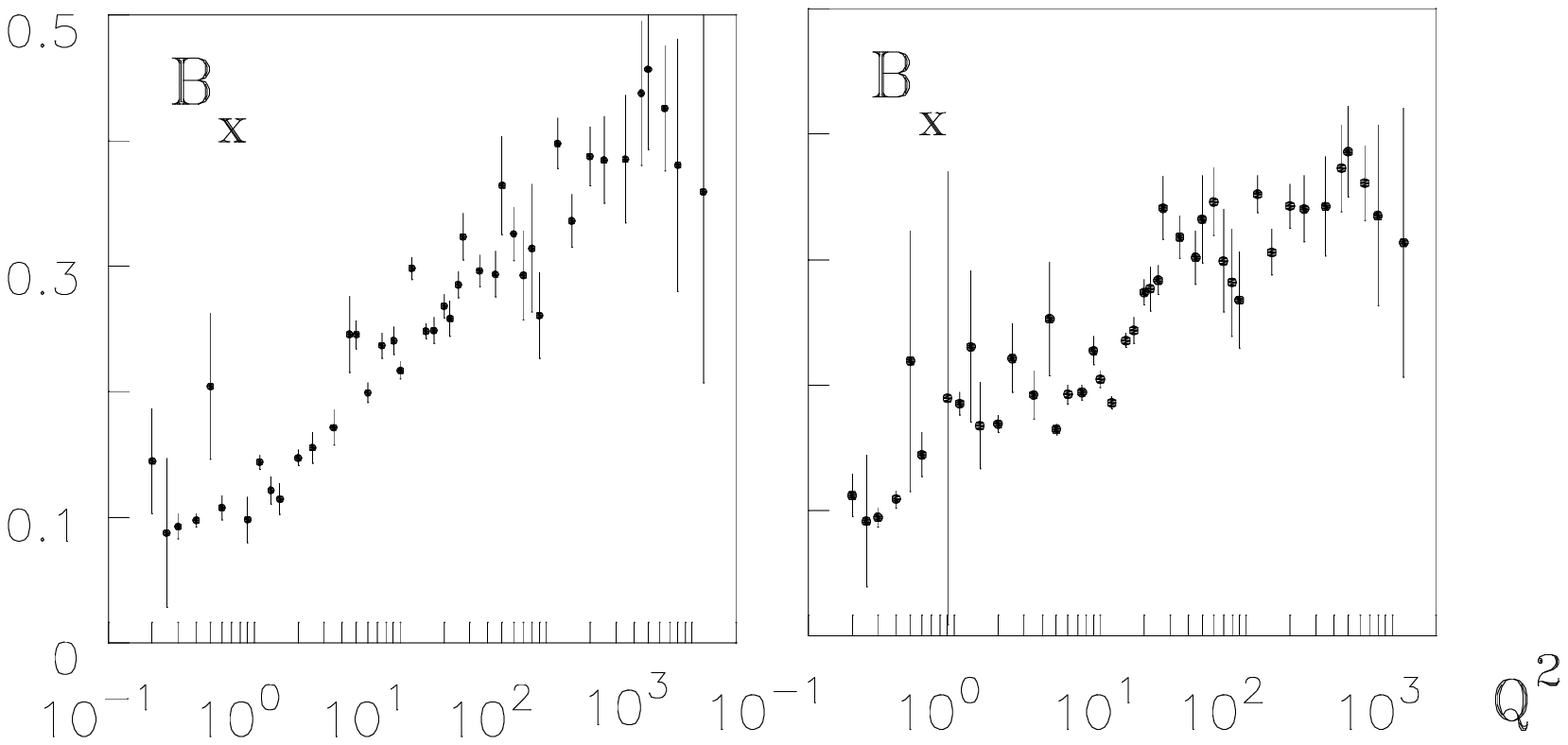}
\end{center}
Fig.~4. $x$ slope $B_x$ versus $Q^2$ and parametrizations (5)~: 
$F_2=a_0({\frac 1x})^{0.418}\ +a_1({\frac 1x)}^{0.0808}$ (left side) and (7)~:
$F_2^{S,0}=b_0+b_2\ell n^2({\frac 1x})$ (right side).

Asymptotically, as $x\rightarrow 0$, the $x$-slope $B_x$
calculated from eqs. (5)-(8) is $Q^2$ independent. However, for
finite values of $x$, in the range of the present experiments, the
$Q^2$ dependence, as it can be seen from Fig.~1,4 is still
essential.
\bigskip

\section{Conclusions}

Our comparative analysis shows that several competing parametrizations for the
small $x$ structure function exist, providing equally good fits to
the data. The $Q^2$ dependent intercept in (1) may be considered
as an "effective" one, reflecting the contribution from two Pomerons
in (2).  Notice that logarithmic parametrization (3) conserving
the unitarity bounds is equally efficient.

The fits show also some evidence that the rise of
the singlet component of the structure functions with $1/x$ moderates
as $Q^2$ increases, the turning point being around $Q^2=200$ GeV$^2$,
whenafter  $F_2(x,Q^2)$ decelerates monotonically. 
Such a slow-down (deceleration) of the rate of increase was anticipated 
already in~\cite{RW}. Later, it was confirmed and discussed in the 
framework of a model~\cite{DJP} interpolating (combining) between Regge 
behaviour and the high $Q^2$ asymptotics of the GLAP evolution equation. 
It was discussed also in ~\cite{DLM} with a traditional
Regge-type model with a $Q^2$ independent Pomeron intercept.

Apart from the turn-over in the $x$ slope $B_x$, the "softening" of the
singlet SF towards highest $Q^2$ may be visualized also from the behaviour of
the fitted $Q^2$ dependent coefficients, namely $a_1$ in Fig. 2,
and $b_2$ in Fig. 3.

Here, we only mention that the origin of the phenomenon - if
confirmed - is either the increasing role of shadowing as $Q^2$
increases, restoring the Froissart bound, or the revelation of a
contribution different from the "perturbative" Pomeron (whose role
was believed to increase with increasing virtuality $Q^2$), or a
combination of the two.

As discussed in~\cite{schof}, the concavity of the slope $B_x$ with respect
to  $Q^2$ is another important quantity, indicative of the
path of evolution (GLAP or BFKL).

Both (BFKL and GLAP) evolution equations are known to be the
theoretical bases of the small $x$ behaviour of the structure functions.
While the perturbative solution is well known for the GLAP equation, its
convergence for the BFKL equation is still debated.
Approximate solution of both and the relevant path in the $x-Q^2$
plane may be revealed both from phenomenological models and from fits
to the data.

Present data may
already reveal~\cite{future} the actual path, but it should be
remembered that the highest measured values of $Q^2$ do not reach the
smallest $x$, so further
measurement at possibly smallest $x$ and highest $Q^2$ are eagerly
awaited.


\begin{thebibliography}{99}


\bibitem{CDD}  A. M. Cooper-Sarkar, R. C. E. Devenish, A. De Roeck, Int. J.
Mod. Phys. ABDI {\bf 13} (1998) 3385


\bibitem{dl}  A. Donnachie, P.V. Landshoff, Phys. Lett. B 437 (1998) 408


\bibitem{cdl}  J. R. Cudell, A. Donnachie, P.V. Landshoff, e-Print Archives:
hep-ph/9901222, 1999


\bibitem{BGJPP}  M. Bertini, M. Giffon, E. Predazzi, Phys. Lett.
B {\bf 349} (1995) 561;
M. Bertini {\it et al.}, Rivista Nuovo Cim. {\bf 19} (1996) 1;
P. Desgrolard, L.Jenkovszky, F. Paccanoni: Can the Pomeron
(diffraction) be "soft" or "hard"?,  
"Hadrons-98", Int. Conf. on Strong Interaction at High Energies,
Parthenit, Crimea (june 1998).
In Proceedings, edited by L. Jenkovszky (1998) p.78.


\bibitem{log}  L. Jenkovszky, F. Paccanoni, E. Predazzi, 
Int. Conf. on
Diffractive and Elastic Scattering, La Biodoba, Isola d'Elba, Italy (may 1991).
Nucl. Phys. B (Proc. Supp.) {\bf 25} (1992) 86;
P. Desgrolard {\it et al.}, Phys. Lett. B {\bf 309} (1993) 191;
M. Bertini {\it et al.}, in "Strong Interactions at Long Distances",
edited by L. Jenkovszky (Hadronic Press Inc., Palm Harbor FL USA, 1995) p.235;
D. Haidt, W. Buchmuller, e-Print Archives: hep-ph/9605428, 1996;
O. Schildknecht, H. Spiesberger, e-Print Archives: hep-ph/9707447, 1997


\bibitem{JLP}  L. Jenkovszky, A. Lengyel, F. Paccanoni, Nuovo Cimento A
{\bf 111} (1998) 551


\bibitem{BDI} M. Bertini, P. Desgrolard, Yu. Ilyin, Int. J. Phys.
{\bf 1} (1995) 45.


\bibitem{kaidalov}  A. Capella {\it et al.}, Phys. Lett. B {\bf 337} (1994) 358


\bibitem{HERA1}  M. Derrick {\it et al}, ZEUS collaboration, Z. Phys. C
{\bf 63} (1994) 391;
S. Aid {\it et al}, H1 collaboration, Z. Phys. C {\bf 69} (1995) 27;
M. Derrick {\it et al}, ZEUS collaboration, Z. Phys. C {\bf 72} (1996) 399;
J.Ahmed {\it et al}, H1 collaboration, Nucl. Phys. B {\bf 470} (1996) 3;
M.R. Adams {\it et al}, E665 collaboration, Phys. Rev. D {\bf 54} (1996) 3006;
C.Adolf {\it et al}, H1 collaboration, Nucl. Phys. B {\bf 497} (1997) 3;
J. Breitweg {\it et al}, ZEUS collaboration, Phys. Lett. B {\bf 407} (1997)
432;
J. Breitweg {\it et al}, ZEUS collaboration, ZEUS Results on the Measurement
Phenomenology of F2 at Low x and Low $Q^2$, DESY-98-121, e-Print Archives:
hep-ex/9809005 (1998)


\bibitem{NMC}  A.C.Benvenuti {\it et al}, BCDMS collaboration, Phys. Lett. B
{\bf 223} (1989) 485;
M. Arneodo {\it et al}, NMC collaboration, SLAC-PUB\ 317 (1990);
L.W. Whitlow {\it et al}, Phys. Lett. B {\bf 282} (1992) 475;
M. Arneodo {\it et al}, NMC collaboration, Nucl. Phys. B {\bf 483} (1995) 3


\bibitem{IL}  Yu. Iljin, A. Lengyel {\it Alternating series fit to the total
cross sections}, ITP-93-25E, Kiev (1993)


\bibitem{schof} H. Navelet, R. Peschanski, S.Wallon, Mod. Phys. Lett. A {\bf 9}
(1994) 3393;
L. Schoeffel, Univ. Paris XI / Orsay,
Dissertation thesis (12-1997), http://www-h1.desy.de/ 


\bibitem{RW}
A. De Roeck, E.A. De Wolf, Phys. Lett. B {\bf 388} (1996) 188


\bibitem{DJP} P. Desgrolard, L. Jenkovszky, F. Paccanoni, Eur. Phys. J. C
{\bf 7} (1999) 263


\bibitem{DLM} P. Desgrolard, A. Lengyel, E. Martynov, Eur. Phys. J. C
{\bf 7} (1999) 655


\bibitem{future}
P. Desgrolard {\it et al.}, paper in progress.

\end{thebibliography}
\end{document}